\documentclass[aps,prl,twocolumn,aps,floatfix,showpacs,showkeys,superscriptaddress]{revtex4-2}

\usepackage{amssymb}
\usepackage{amsbsy}
\usepackage{amsthm}
\usepackage{amsmath}
\usepackage{breqn}
\usepackage{bm}
\usepackage{cancel}
\usepackage{braket}
\usepackage{color}
\usepackage{dcolumn}
\usepackage{epsfig}
\usepackage[T1]{fontenc}
\usepackage{graphicx}
\usepackage[colorlinks]{hyperref}
\usepackage{mathtools}
\usepackage{placeins}
\usepackage{xcolor}
\usepackage{cleveref}
\usepackage{subfigure}

\begin{document}
	
\title{Quantum Liang Information Flow Probe of Causality Across Critical Points}

\author{Roopayan Ghosh}
\thanks{ucaprgh@ucl.ac.uk}
\affiliation{Department of Physics and Astronomy University College London$,$ Gower Street$,$ London$,$ WC1E 6BT$,$ United Kingdom}
\author{Bin Yi}
\affiliation{Institute of Fundamental and Frontier Sciences$,$ University of Electronic Science and Technology of China$,$ Chengdu 610051$,$ China}
\affiliation{Department of Physics and Astronomy University College London$,$ Gower Street$,$ London$,$ WC1E 6BT$,$ United Kingdom}

\author{ Sougato Bose}
\affiliation{Department of Physics and Astronomy University College London$,$ Gower Street$,$ London$,$ WC1E 6BT$,$ United Kingdom}



	\date{\today}
\begin{abstract}
Investigating causation in the quantum domain is crucial. Despite numerous studies of correlations in quantum many-body systems, causation, which is very distinct from correlations, has hardly been studied. We address this by demonstrating the efficacy of the newly established causation measure, quantum Liang information flow, in quantifying causality across phase diagrams of quantum many-body systems. We focus on quantum criticality, which are  highly non-classical points. We extract causation behavior across a spectrum-wide critical point and a ground state second-order phase transition in both integrable and non-integrable systems. Across criticality, each case exhibits distinct hallmarks, different from correlation measures. We also deduce that quantum causation qualitatively follows the quasiparticle picture of information propagation in integrable systems but exhibits enhanced quantum non-locality near criticality. At times significantly larger than the spatial separation, it extracts additional features from the equilibrium wavefunction, leading to a peak just before the critical point for near boundary sites.
\end{abstract}

	\maketitle
\paragraph{\textbf{Introduction:}}
The study of dynamics of quantum many-body systems typically involve the time evolution of correlation functions and the spreading of entanglement \cite{PhysRevA.69.022304,RevModPhys.83.863,PhysRevA.84.052316,doi:10.1142,PhysRevB.105.144202}. Even the study of lightcones, operator spreading\cite{PhysRevE.56.5174,PhysRevE.64.036207,PhysRevLett.89.060402,PhysRevE.70.016217,PhysRevD.106.046007}, Out-of-time-ordered-Correlator (OTOC)\cite{Maldacena2016,PhysRevB.96.060301,PRXQuantum.2.020339,PhysRevB.99.054205,PhysRevE.107.064207,PhysRevLett.124.140602}, quantum chaos measures \cite{PhysRevX.10.041017,lim2024defining} are essentially correlation studies. On the other hand, quantifying causation dynamics in the quantum realm, unlike classical theories, has been challenging \cite{Brukner2014}. In fact, the adage "correlation does not always imply causation" holds true even in the context of quantum mechanics,  and the most commonly studied measures can at best detect a causal connection but not {\em quantify} the amount by which subsystems influence each other. We address this limitation in our work.

 Model Hamiltonians of quantum chains provide the microcosm of quantum effects prevalent in our universe. Hence, it is of paramount importance to find a measure of causation in these systems, which is easily measurable and can be appropriately connected to the physical intuition born from classical systems, yet demonstrating distinct quantum signatures. This has led to efforts in which quantum causation is extracted from quantum correlations, drawing inspiration from the Liang-Kleeman analysis used in classical systems\cite{, PhysRevLett.95.244101,san2016information, liang2018causation}, using von Neumann entropy\cite{yi2022quantum}. The quantum version of Liang information flow (where the causal influence is defined as a property of interaction between two subsystems) offers distinct advantages over observable-based correlation measures that purportedly detect similar behavior. First, its use of information-theoretic tools makes it more universal. Second, it is intuitively connected to the classical picture and is easy to implement in experimental setups, as its simplest version requires only single-site measurements. However, its true test lies in whether it can properly quantify non-classical phenomena. Critical regions in quantum many-body spin systems provide the perfect playground for this evaluation, which is where we focus our efforts.  

\begin{figure}
    \centering
    \includegraphics[width=0.6 \columnwidth]{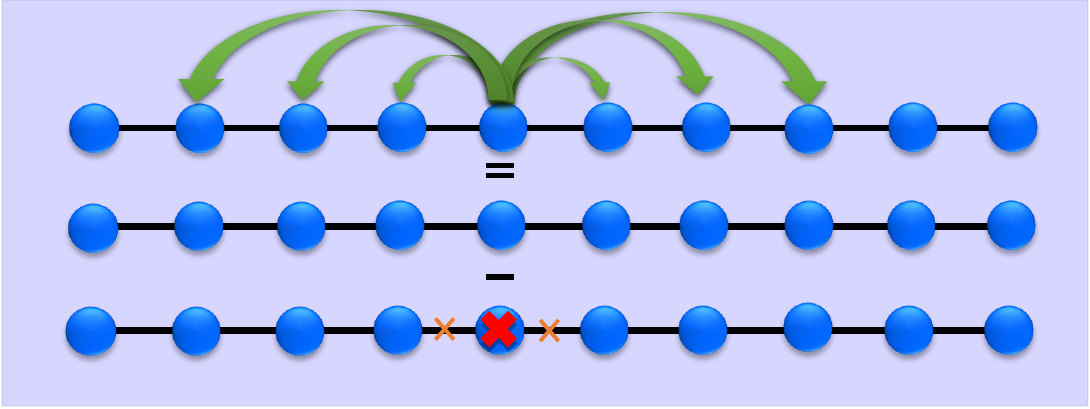}\\
    
   \includegraphics[width=0.6 \columnwidth]{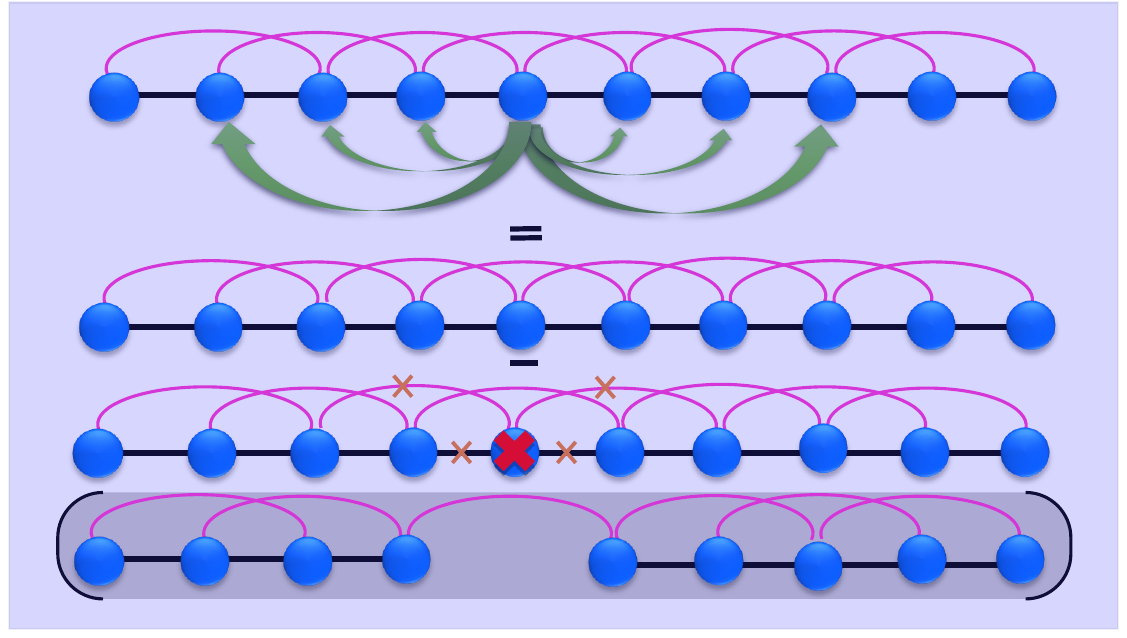}
    \caption{Schematic of Liang information flow (green arrows) between a chosen site and others, computed by comparing a information flow in a normal lattice to one with a frozen site (cross). The top diagram shows nearest-neighbor coupling, while the bottom includes next-nearest-neighbor coupling. }
    \label{fig:schematic}
\end{figure}
In this work, we show that distinct signatures of quantum  criticality are seen in quantum causation quantified by quantum Liang information flow. The key difference compared to just measuring entanglement entropy is that an intervention is applied, which allows us to measure quantum causation, not just correlations. To demonstrate, we conduct simulations of quench dynamics \cite{annurev} using model Hamiltonians. Previous research has  explored the time evolution of correlation functions for quenches across critical points in quantum many-body systems\cite{PhysRevA.69.053616,Basu2012}, leveraging non-analyticities to detect phase transitions\cite{PhysRevX.11.031062,PhysRevB.107.094432}, but these do not demonstrate causality. Some studies involving OTOC are also available in the literature\cite{PhysRevB.96.054503,Sun:2020nfk,PhysRevLett.121.016801},  but their behaviour frequently depends on the choice of operator due to possible conservation laws and they do not {\em quantify} causation. Our work suffers from no such drawbacks.


\paragraph{\textbf {Definitions:}}
\label{sec:def}
For density operator $\rho$, the von Neumann entropy $S(\rho)$ is defined as $
S(\rho)=-{\rm Tr}[\rho \log\rho]$.
Consider a system $S$ with state $\rho$ evolving under the unitary $U(t)$ generated by Hamiltonian $H$. Let the reduced density matrices of subsystems $A$, $B$, and $AB$ be $\rho_A$, $\rho_B$, and $\rho_{AB}$, respectively. Following Ref.~\cite{yi2022quantum}, the quantum Liang information flow is  
\begin{equation}
    T_{B\rightarrow A}=\frac{dS(\rho_A)}{dt}-\frac{dS(\rho_{A\not{B}})}{dt}, \label{rflow1}
\end{equation}
where $\rho_{A\not{B}}$ is the density matrix of $A$ with $B$ frozen. While in Ref.~\cite{yi2022quantum} the derivative was used to denote rate of change of von-Neumann entropy, for our purpose, the r.h.s. can be considered to symbolically represent the Liang information flow at an infinitesimal time interval. If the system with frozen $B$ evolves for time $t$, then the cumulative Liang information flow is  
\begin{equation}
    \mathbb{T}_d=S(\rho_A,t)-S(\rho_{A\not{B}},t),
    \label{cumlative}
\end{equation}
where $d$ is the distance between $A$ and $B$. This quantifies the \textit{total} influence (Liang information flow) from $B$ to $A$ over time by tracking the change in $A$’s entanglement with the system excluding $B$. Freezing a site effectively switches off certain Hamiltonian couplings, making Eq.~\eqref{cumlative} a measure of the `local quench'~\cite{Fukuhara2013,PhysRevE.89.062110} effect on $A$. The setup is schematically illustrated in Fig.~\ref{fig:schematic}, see also App.~\ref{app:appA}. 

\paragraph{\textbf{Localization transition in Aubry-Andre-Harper model:}}
\label{sec:and}
\begin{figure}
    \centering
    \includegraphics[width=0.45 \columnwidth]{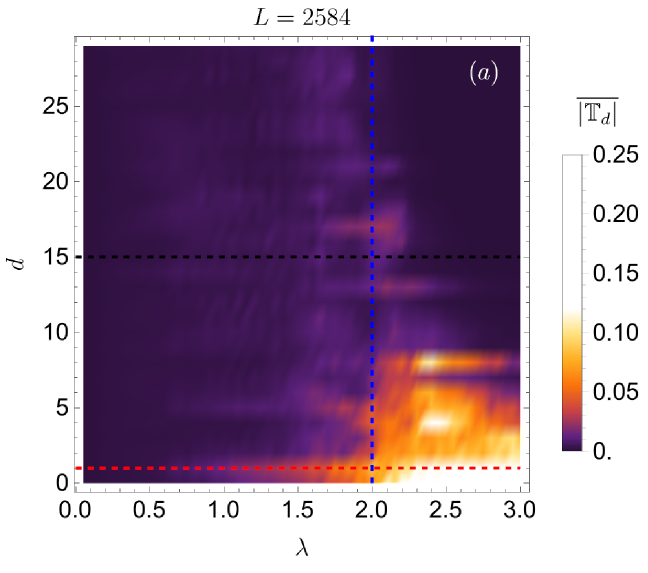}
    \includegraphics[width=0.45 \columnwidth]{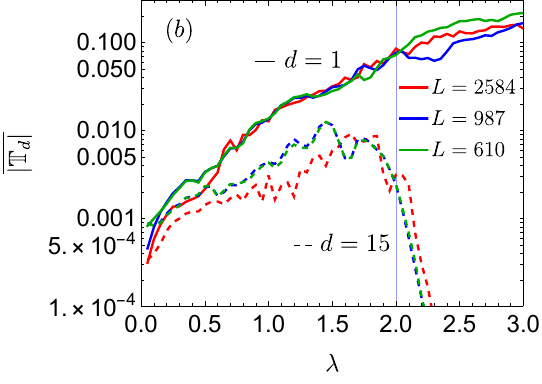}
    \caption{Cumulative Liang information flow for the model in Eq.~\eqref{eq:aa}. (a) Late-time cumulative flow to a site $d$ sites away from the frozen site, with $|T_d|$ averaged over $t \sim 10^2$ to smooth the plot. The blue dashed line marks the localization phase transition at $\lambda = 2$. (b) Cross section of (a) at the red dashed line, with similar data for other system sizes in different colors. The blue line marks the critical point. The frozen site position is the closest smaller Fibonacci number +1 to $L$, e.g., for $L=2584$, $i=1598$, to minimize finite size effects.}
    \label{fig:aafree}
\end{figure}
 We first examine Liang information flow in the 1D  Hamiltonian:
\begin{equation}
    H=\sum_{j=1}^{L-1} (\sigma^x_j \sigma^x_{j+1}+\sigma^y_j \sigma^y_{j+1})+\frac{1}{2}\sum_{j=1}^L\mathcal{B}_j\sigma^z_j.
    \label{eq:aa}
\end{equation}
where $\sigma^{x,y,z}$ are Pauli-spin operators, $\mathcal{B}_j=\lambda \cos( 2 \pi \beta j)$, and $\beta=\frac{\sqrt{5}-1}{2}$, the inverse golden ratio. The system size $L$ is chosen as a Fibonacci number to minimize finite size effects\cite{PhysRevLett.51.1198}. This model undergoes a localization-delocalization transition at $\lambda=2$ across its eigenspectrum~\cite{aubry1980analyticity}.

In Fig.~\ref{fig:aafree}, we analyze the causal influence of a selected site across the transition using cumulative Liang information flow $\mathbb{T}_d$ between sites at distance $d$, averaged beyond the transient growth ($t \sim 10^2$).
\begin{figure}
    \centering
    \includegraphics[width=0.46 \columnwidth]{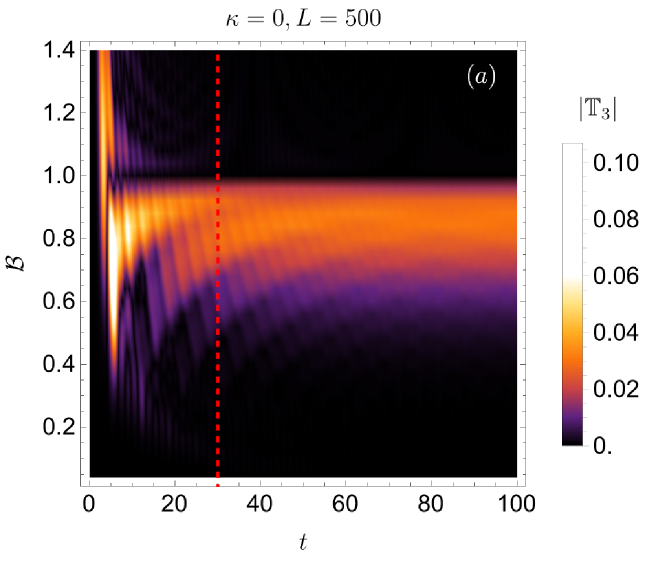}
    \includegraphics[width=0.46 \columnwidth]{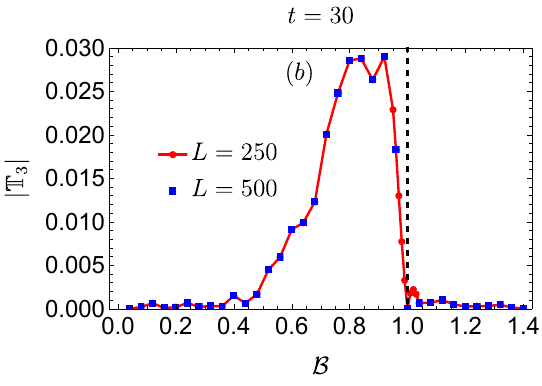}
    \includegraphics[width=0.46 \columnwidth]{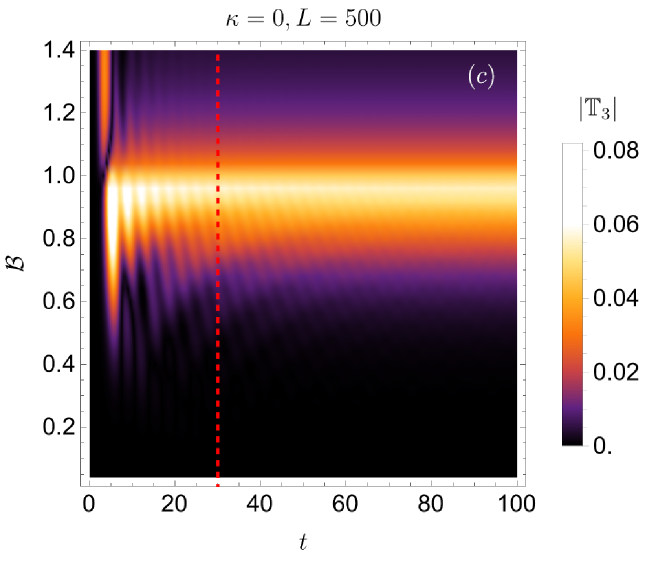}
    \includegraphics[width=0.46 \columnwidth]{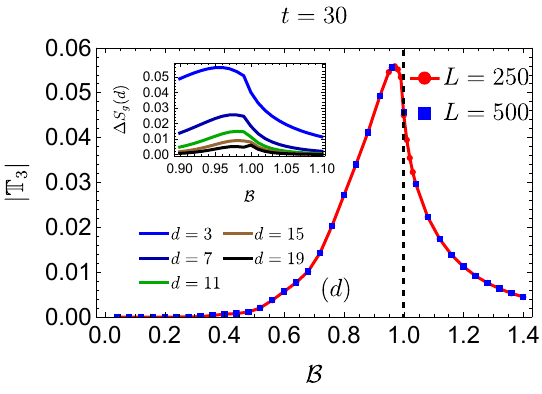}  
    \includegraphics[width=0.46 \columnwidth]{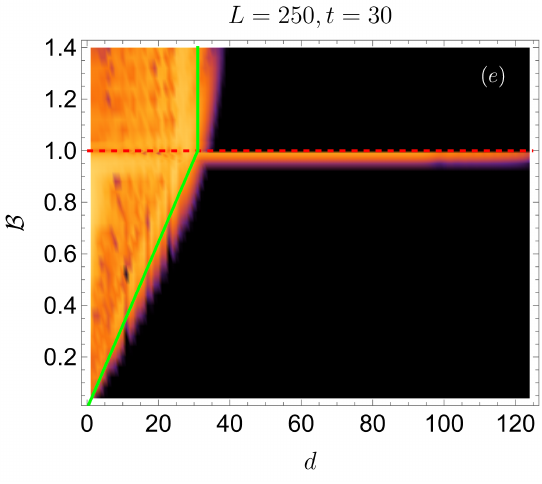}  
    \includegraphics[width=0.46 \columnwidth]{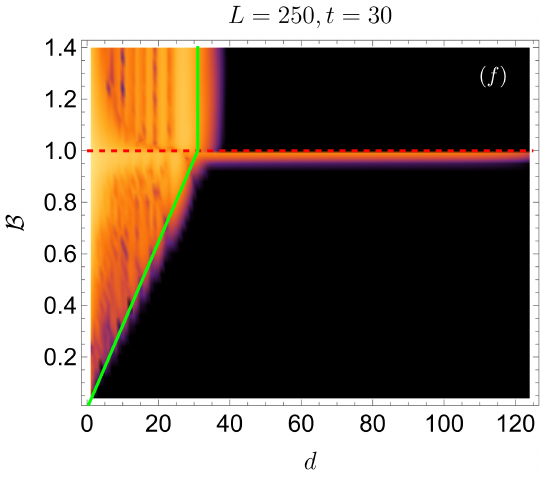}  \\ \vspace{-0.2 in}
    \includegraphics[width=0.4 \columnwidth]{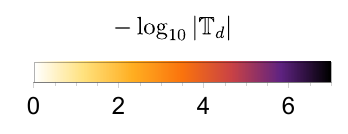}  
    \caption{Variation of cumulative Liang information flow ($|\mathbb{T}_d|$) for quenches to different $\mathcal{B}$ fields at $\kappa=0$. (a), (b): Initial state is the ground state of $\mathcal{B}=0.01$. (c), (d): Initial state is the ground state of the respective $\mathcal{B}$. The inset of (d) shows $\Delta S_g(d)$, the entanglement difference at distance $d$ from the frozen site, between normal and frozen-site ground states, as a function of $\mathcal{B}$. (a) shows $|\mathbb{T}_3|$ at three sites from the frozen site; red dashed lines mark the cross section in (b), where the black line indicates the critical point. (e), (f): Spatial profiles of Liang information flow for quenches at $L=250$ at a chosen time. (e) follows (a)’s setup, and (f) follows (c)’s. The red dashed line marks $\mathcal{B}_c=1$, and green lines show the fastest quasiparticle’s reach. }
    \label{fig:causalitymaximakappa0}
\end{figure}
In Fig.~\ref{fig:aafree}(a), we observe that in the delocalized regime, there is equitable Liang information flow from the frozen site to others. This corroborates the unrestricted transport expected in this regime as a result of spatially extensive single-particle eigenfunctions. This indicates that in large delocalized and ergodic systems, removing a few qubits, no matter their location in the chain, has no effect on dynamics. However, as $\lambda$ nears the critical point, causation effects increase for nearby sites due to the restriction of information propagation beyond the localization length. Thus, local effects dominate a site’s evolution. Since the phase transition spans the eigenspectrum, the initial state’s energy has minimal impact. Therefore, our choice of the Neel state as the initial state gives similar qualitative results to other typical states.~\footnote{See supplementary material including Ref.~\cite{PhysRevB.103.024202} therein.}).

In Fig.~\ref{fig:aafree}(b), we focus on two specific scenarios: (i) $d=1$, representing a nearby site to the frozen site, and (ii) $d=15$, representing a site at a distance greater than the localization length in the localized regime. For the site at $d=1$, we observe the anticipated behavior described in the preceding paragraph as $\lambda$ increases. However for distant sites the causation flow becomes intriguing near criticality. For small $\lambda$, the behavior mirrors that of nearby sites i.e. there is a gradual increase in cumulative Liang information flow with $\lambda$ until approximately $\lambda \sim 1.5$. This is followed by a plateau before the expected sharp decline in the localized regime. This indicates a peak in causality occurs for an innocuous parameter, $\lambda<2$, for distant sites. It also shows a slow drift towards $\lambda \sim 1.5$ for larger values of $d$ as seen in Fig.~\ref{fig:aafree}(a), from the faint violet regions. The larger causation values indicate an already inequitable flow of information between different sites in the said parameter regime. This in turn signifies traces of localization in the parts of the system acting as a herald to the onset of localization across the spectrum for $\lambda=2$. A naive finite size analysis in Fig.~\ref{fig:aafree}(b) suggests this effect is largely independent of system size. Furthermore, for $\lambda>2$ exponentially localized wavefunctions result in exponentially small information leakage beyond the localization length, whose signature is given by a rapid but continuous decrease in Liang information flow with $d$.

\paragraph{\textbf{Ground state phase transitions in Ising models:}} 
\label{sec:annni}

Next, we direct our focus to the Anisotropic Next Nearest Neighbor Ising (ANNNI) model with open boundary conditions (obc). This model includes a next-nearest-neighbor coupling, is non-integrable, and lacks $U(1)$ symmetry present in the previous example. In certain parameter regimes, the ground state of this model undergoes a ferromagnetic to paramagnetic Ising phase transition upon tuning the transverse magnetization strength. In the following analysis, we will investigate the influence of the middle site of the chain on other chosen sites across different parameter regimes. Note that choosing any other site in the bulk gives qualitatively same results. 

The Hamiltonian of ANNNI chain is given by,
\begin{equation}
	H_{L}=-\sum_{j}^{L-1}\sigma_j^z\sigma^z_{j+1}+\kappa\sum_{j}^{L-2}\sigma_j^z\sigma_{j+2}^z	-\mathcal{B}\sum_j^L\sigma_j^x,
	\label{annni chain}
\end{equation}
 where $\kappa$ represents the strength of the next-nearest neighbor term and  $\mathcal{B}$ is the magnitude of magnetic field applied along the  transverse axis.
 
 At fixed value of $0\le\kappa<0.5$, the ground state of the ANNNI model undergoes quantum phase transition from the ferromagnetic to the paramagnetic phase when transverse field $\mathcal{B}>0$ exceeds a critical value $\mathcal{B}_c$ \cite{SELKE1988213,PhysRevE.75.021105,Suzuki2013,DAllen_2001,Nagy_2011,PhysRevB.66.064413,PhysRevB.73.052402}, and the critical parameters $\kappa_c$ and $\mathcal{B}_c$ are satisfies:
 \begin{equation}
1-2\kappa_c=\mathcal{B}_c-\mathcal{B}_c^2\frac{\kappa_c}{2-2\kappa_c}
\label{eq:criticalB}
 \end{equation}

While this phase transition occurs in equilibrium, previous studies \cite{10.21468/SciPostPhys.15.1.032,PhysRevX.11.031062,PhysRevB.107.094432,PhysRevB.107.L121113,PhysRevE.74.031123} show that its signatures can appear in non-equilibrium quantum quenches, motivating us to explore these within Liang information flow.

Since the model in Eq.~\eqref{annni chain} is non-integrable for generic parameter values we resort to numerical simulations (TDVP and DMRG\cite{itensor-r0.3}) for our results. However, we first explore the integrable limit $\kappa=0$, before discussing the results for $\kappa >0$. 
\paragraph{$\kappa=0$:}
For $\kappa=0$, we apply a canonical transformation ($\sigma^x \rightarrow \sigma^z, \sigma^z \rightarrow -\sigma^x$) and a Jordan-Wigner transformation to map $H_L$ to spinless fermions, yielding $\mathcal{B}_c=1$ in the thermodynamic limit. The one-site density matrix required for computation is $\rho_j = \frac{\mathbb{I} + \langle \sigma^z_j \rangle \sigma^z}{2}$ \cite{PhysRevA.66.032110}. The time evolution of $\langle \sigma^z \rangle$ can be found semi-analytically due to the Hamiltonian's quadratic form in fermionic operators.~\footnote{See supplementary material and Ref.~\cite{IngoPeschel_2003,Peschel_2009} therein}.

 In Fig.~\ref{fig:causalitymaximakappa0}(a) we show the flow of Liang information for $d=3$, when the initial state is the ground state at $\mathcal{B}=0.01$ and we quench to different values of $\mathcal{B}$. When $\mathcal{B}$ is small, the initial state in this regime has large overlap with the ground state thus limiting non trivial evolution and Liang information flow. The flow gradually increases as we increase $\mathcal{B}$ towards $\mathcal{B}_c$ due to increased quantum fluctuations introduced by the $\sigma^x$ in the dynamics. However, if $\mathcal{B}$ is too large (deep inside paramagnetic phase), evolution becomes dominated by the local $\sigma^x$ term and we notice the expected qualitative behaviour of small Liang information flow in Fig.~\ref{fig:causalitymaximakappa0}(b) for $\mathcal{B}>1$. But the peak of quantum causation appears to be shifted from the critical $\mathcal{B}_c=1$, where one would expect the quantum effects in the evolution the most. One possible cause for this is our choice of initial state having significant overlaps with several high energy states in this regime. To verify this, we study another initial state: the ground state of the corresponding $\mathcal{B}$ where causality is being computed, the results of which is shown in Figs.~\ref{fig:causalitymaximakappa0} (c) and (d), which gives us almost the expected features with the peak occurring closer to $\mathcal{B}_c$. The reason for the remaining skewness towards the ferromagnetic phase is the greater influence of nearby neighbours in the dynamics due to dominance of a nearest-neighbour term, which is seen in the maxima of $|\mathbb{T}_d|$ when $t \gg d $. The effect becomes more prominent for non-ground initial states. Furthermore, from the well-known quasiparticle picture of information propagation for local quenches~\cite{Calabrese_2007} which holds for Fig.~\ref{fig:causalitymaximakappa0}(d), we deduce that for $t\gg d$, $|\mathbb{T}_d|$ is effectively the difference of $S(\rho_A)$ for site $A$ at a distance $d$ from the frozen site, for the ground states of the unfrozen and frozen system. We call this quantity $\Delta S_g(d)$ and plot it in the inset to show the resemblance. As evident from the plot, this feature is prominent near the edge (small $d$) where the flow is strongest and is embedded in the ground state of the model. This concludes our explanation of the peak-before-criticality phenomena.

Finally, we plot the spatial profile of Liang information flow at time $t=30$ in Fig.~\ref{fig:causalitymaximakappa0}(e) and (f). 
We make two key observations in these plots. First, the causation shows a spatial envelope. This is again qualitatively consistent with the picture: quasiparticles carry information in the system~\cite{Calabrese_2007, 10.21468/SciPostPhys.11.3.055}. To corroborate, we show the distance covered by the fastest quasiparticle for $t=30$ by green line, whose velocity is  ${\rm Min}(1,\mathcal{B})$. The small deviations can be attributed to the choice of initial state and finite sizes. Secondly, regardless of the initial state chosen, there is a small but highly non-local causation near/at the critical point which coincides with the divergence of correlation length of the ground state. This behaviour is caused by the participation of a edge localized eigenmode~\footnote{Ferromagnetic phase of $\kappa=0$ for obc hosts this mode, see Ref.~\cite{10.21468/SciPostPhys.3.3.020} and supplementary material} with diverging localization length near criticality in the dynamics. This effect is beyond the quasiparticle picture, as there is no propagation of plane wave (sinusoidal) eigenmodes, yet an extended eigenstate with long range entanglement participates in dynamics, which exerts a quantum non-local causation upon freezing a site. This is a rare example of beyond-quasiparticle quantum non-local behaviour in spin systems. Immediately after critical point (which can be shown to be at $\mathcal{B}_c \sim L/(L+1)$ due to finite-size effect), this mode changes to a plane wave and the non locality vanishes~\footnote{See supplementary material and references~\cite{PhysRevA.66.032110, 10.21468/SciPostPhys.15.1.032,Calabrese_2007, LIEB1961407,PhysRevLett.67.161,PhysRevB.84.165117} therein for full computation details}. 

\begin{figure}
    \centering
    \includegraphics[width=0.46 \columnwidth]{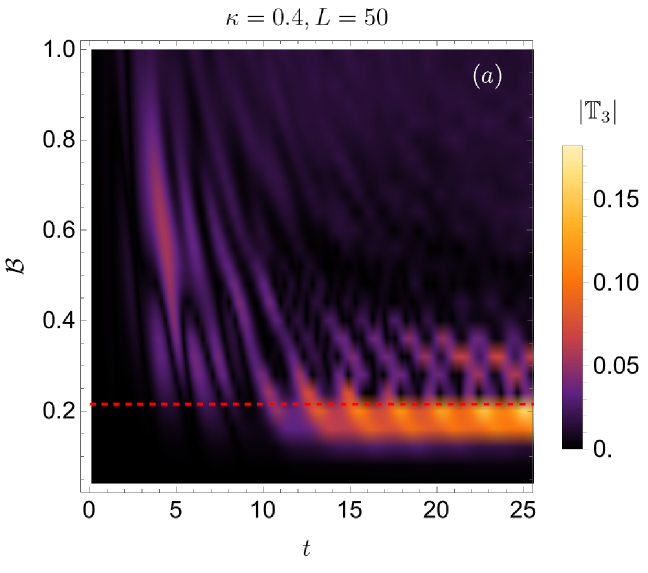}
    \includegraphics[width=0.46 \columnwidth]{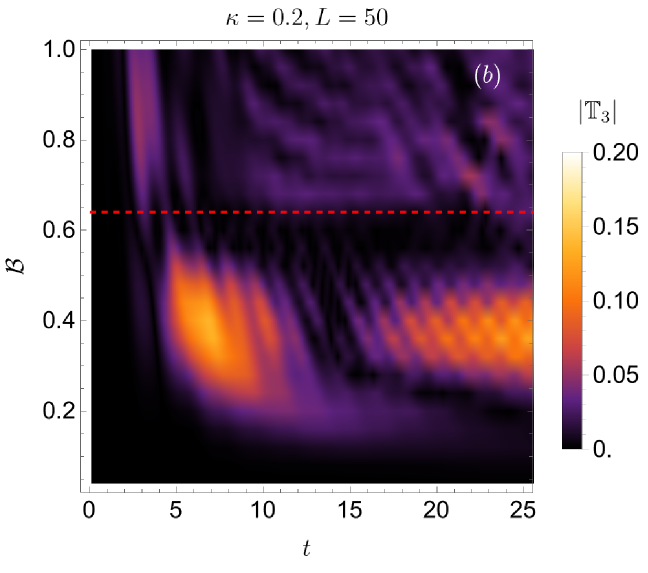}
    \includegraphics[width=0.46 \columnwidth]{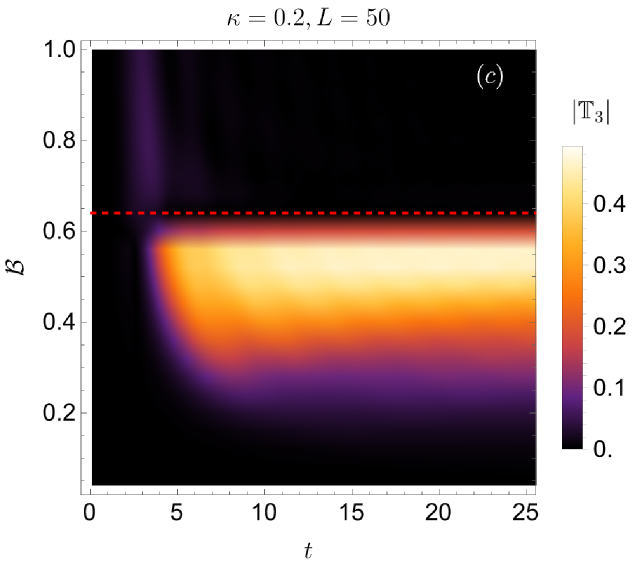}
    \includegraphics[width=0.46 \columnwidth]{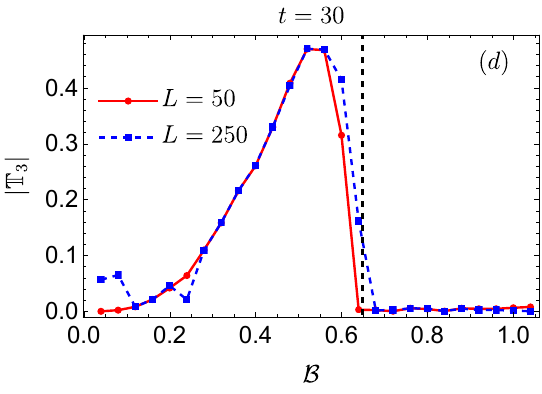}  
    \caption{Variation of absolute Liang information flow for quenches to different $\mathcal{B}$ fields at fixed $\kappa \neq 0$, computed via TDVP for system sizes $L=50, 250$. (a) $|\mathbb{T}_3|$ at three sites from the frozen site for $\kappa=0.4$, starting from the ferromagnetic state $\ket{\downarrow \downarrow \hdots \downarrow}$. (b) Same as (a) for $\kappa=0.2$. (c) $|\mathbb{T}_3|$ for $\kappa=0.2$ starting from the ground state, similar to Fig.\ref{fig:causalitymaximakappa0}~(c). (d) Cross section of (c) at $t=30$ compared with $|\mathbb{T}_3|$ for $L=250$. Red/black dashed lines denote $\mathcal{B}_c$ from Eq.\eqref{eq:criticalB}.}
    \label{fig:causalitymaxima1}
\end{figure}

\paragraph{$\kappa>0$:}
For $\kappa > 0$, performing quenches in $\mathcal{B}$ or $\kappa$ yields similar results~\footnote{See supplementary material}. In Fig.~\ref{fig:causalitymaxima1} we show the features shown by quantum Liang information flow for the non-integrable interacting model. As in the previous section we study two initial states: the ferromagnetic state  $\ket{\downarrow\downarrow\downarrow\hdots\downarrow}$ which is the ground state at $\kappa=\mathcal{B}=0$, and the ground state of $\mathcal{B}$ where information flow is computed. \footnote{A tiny longitudinal field $O(10^{-4})$ is added  to prevent degeneracy for small $\mathcal{B}$.} 

In Fig.~\ref{fig:causalitymaxima1}(a), despite starting from the ferromagnetic initial state with a smaller system size $L=50$ compared to Fig.~\ref{fig:causalitymaximakappa0}, we observe a peak in quantum Liang information flow very close to criticality. The proximity of the peak to the critical point compared to the $\kappa=0$ case is due to the fact that larger $\kappa$ values push the critical point towards $\mathcal{B}=0$. Consequently, even at these system sizes, this initial state maintains sufficient overlap with the ground state of the Hamiltonian before the critical point, thereby exhibiting the expected behavior.

But as we decrease $\kappa$ to $0.2$,  the ferromagnetic state no longer maintains a high overlap with the ground state near the corresponding critical point. Furthermore, the system is now ergodic due to its non-integrable nature, implying the evolution of such a state at the same timescales is no longer restricted to the low energy sector(as can be verified by, for example, growth of entanglement). However, the causation peak still necessarily exists because quantum fluctuations has to maximize somewhere in between the extreme cases $\mathcal{B}\ll 1$ and $\mathcal{B}\gg 1$. But it need not bear any relation to the phase transition point, it just depends on the dynamics. This is verified in Fig.~\ref{fig:causalitymaxima1}(b) which shows the peak much before the critical point denoted by the red dashed line. 

Nonetheless, we recover the `peak near criticality' akin to the non-interacting case, once the initial state is the ground state \footnote{See supplementary material for other parameters}, as shown in Figs.~\ref{fig:causalitymaxima1}(c),(d) for Liang Information flow to nearby sites with $d{=}3.$ In the integrable model, starting from even non-ground initial states we can capture the same features of the ground state transition, as seen from Fig.~\ref{fig:causalitymaximakappa0}(e). In contrast, in the non-integrable case, only the low energy initial states show the correct behaviour. Thus, the maximum causality depends on the competing behaviour of the various Hamiltonian terms on the initial state. 
\paragraph{\textbf{Discussion:}}

In this work, we have unraveled causation behavior in quantum chains using the recently formulated quantum Liang information flow. We established that causality peaks do not align with maxima in correlation, as causality is quantified by the {\em difference} in quantum correlations. We found a democracy of influence in the fully delocalized regime and near-site causation in the localized regime. Furthermore, diverging correlation lengths in ground state second-order phase transitions manifest as non-local quantum causation. We also found that the causation peak for nearby sites occurs slightly towards the ordered side of the transition, acting as a non-equilibrium herald to the equilibrium phase transition.

Quantum Liang information flow probe quantifies the influence of different couplings in a Hamiltonian to induce criticality. While classical Liang-Kleeman information has improved AI simulations by uncovering causal relationships in data~\cite{Tyrovolas2023}, our result shows that its quantum counterpart can characterize phase transitions in spin systems, and utilizing it on random graph structures can be key to leveraging quantum computation for NP-hard problems, for example in tackling bottlenecks of quantum annealing~\cite{RevModPhys.90.015002}. Causality peaks naturally delineate dominant interactions around transitions, thus offering a diagnostic tool for detecting structural traits inducing quantum phase transitions, which can be useful in optimizing schedules for critical state preparation, a challenging task. Possible additional applications include identifying resonant regions in many-body localization~\cite{PhysRevB.105.144203, PhysRevB.105.174205} and tracking participating qubits in quantum reservoir computing~\cite{Dudas2023}, similar to its role in improving classical AI~\cite{Tyrovolas2023}.
The examples provided in this work can be tested in D-Wave architecture, where Transverse Ising models with $O(100)$ sites have been simulated~\cite{PhysRevResearch.2.033369, King2023,vodeb2024stirringfalsevacuuminteracting}, or in trapped ion systems~(see App.~\ref{app:appB}). 
\begin{acknowledgments}
RG thanks K Sengupta, A Das, M Sarkar and A Nico-Katz for discussions. RG and SB acknowledge the UKRI EPSRC grants Nonergodic quantum manipulation EP/R029075/1 and  Many-Body Phases In Continuous-Time Quantum Computation EP/Y004590/1 for support. YB acknowledges support from National Natural Science Foundation of China (Grant No.~12404551) and the China Postdoctoral Science Foundation (Grant No.~2024M750339).
\end{acknowledgments}

\bibliography{ref}
\appendix
\section{Further details about the setup}
\label{app:appA}
In the top three chains of Fig.~\ref{fig:schematic}, we depict situations with only nearest-neighbor couplings. In this scenario, removing a site results in a break in the chain, and the effective evolution then occurs within a smaller chain, and the difference of $S(\rho)$ at the target site between the normal and smaller chain gives the Liang information flow.  With longer-range couplings, as the next-nearest neighbour case shown in the lower half of Fig.~\ref{fig:schematic}, freezing one site does not break the chain, positioning the target site consistently within the bulk. However we must emphasize that while computing causation, the breakage of a chain is not a source of any issues. Just to complete this discussion, we would like to mention that to prevent chain breakage upon freezing site $p$, one must ensure couplings of range $p+1$ in a chain with open boundary conditions. Furthermore, for periodic boundary conditions, there is a connection between sites $1$ and $L$, preventing chain breakage, but transforming the system into an open boundary condition system upon freezing a site. However, we opt for open boundary conditions in this work to avoid this additional complexity. Although it should not significantly differ from the results in this work for unitary evolution, discrepancies may arise in the non-unitary scenario, particularly where potential skin effects emerge, and for topological phase transitions. We plan to explore these differences in a subsequent study.
\section{Experimental protocol}
\label{app:appB}
To compute the Liang information flow to a specific distance, one needs to follow a two-step procedure:
\begin{enumerate}
    \item To obtain the results when starting from the ground state of the Ising model, one needs to first initialize the model in its ground state and then select any site, preferably away from the edges of the chain. One then needs to perform single-site tomography to obtain information about the reduced density matrix of the problem. For the integrable Transverse Ising chain, this reduces to $\sigma^z$ measurements at the specific site. On the other hand, to obtain the results starting from the ferromagnetic initial state, one needs to initialize the system in this state and allow it to evolve under the corresponding Hamiltonians for time $t$. Then, one needs to perform single-site tomographic measurements to obtain the reduced density matrix and thus the von Neumann entropy. Note that the reduced density matrix for any general non-integrable for a single site can be reconstructed by three Pauli measurements using the Stokes parameters~\cite{PhysRevA.64.052312}, which can be the protocol for tomography here.
    \item In the second step, one needs to reinitialize the system to either the ground state of the same Hamiltonian as the previous step or the ferromagnetic initial state, depending on which result one intends to obtain. Then, according to the required distance $d$, one should suddenly decouple the site (by turning off the relevant couplings or applying a large local magnetic field) at a distance $d$ from the previously measured site and allow the system to evolve freely. Again, one needs to perform tomographic measurements on the same site as before, after the same time $t$. This would give us the entropy with a site frozen. The difference between these two values yields the Liang information.
\end{enumerate}

\begin{figure}
    \centering
    \includegraphics[width=0.45\columnwidth]{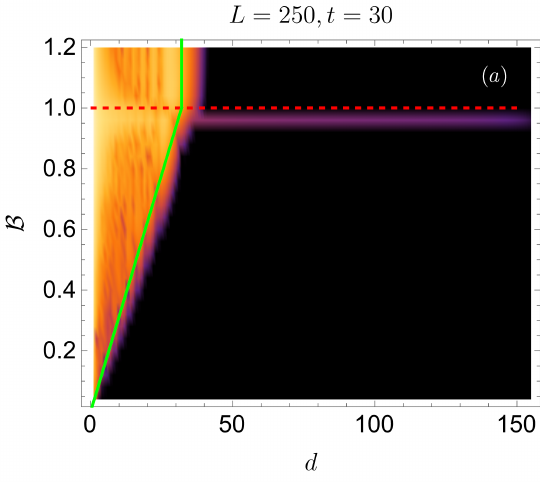}
    \includegraphics[width=0.45\columnwidth]{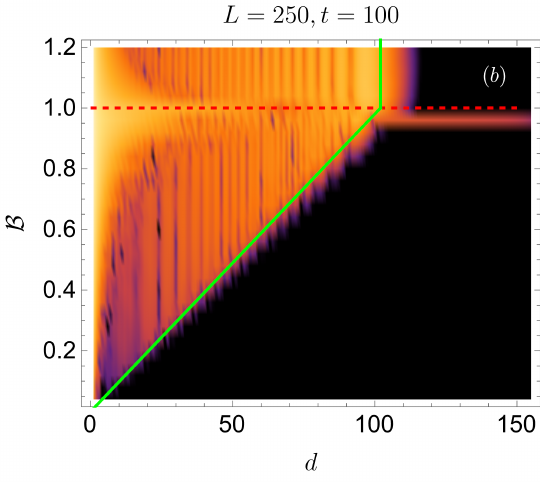}
    \includegraphics[width=0.45\columnwidth]{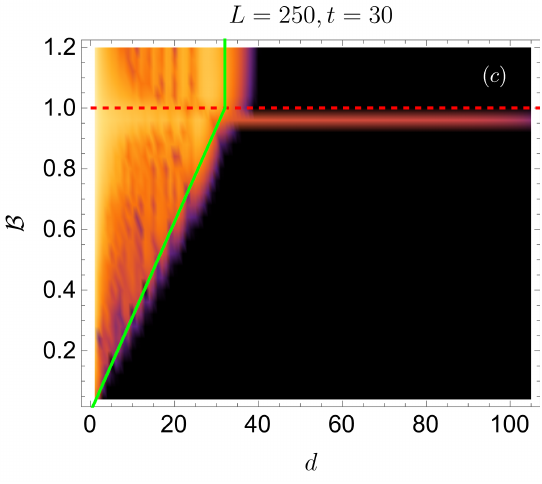}
    \includegraphics[width=0.45\columnwidth]{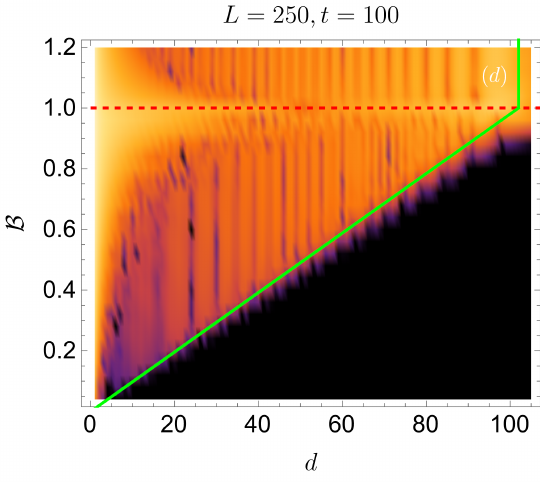}
    \caption{(a) and (b) denotes the spatial profile for Liang information flow for $t=30$ and $t=100$ respectively when site $L/2-30$ is frozen for $L=250$. (c) (d) denotes the corresponding plots when site $L/2+20$ is frozen. There is no qualitative difference in the results. Note that the points on the y-axis are taken $0.04$ apart hence the small gap between there dashed lines which denote $\mathcal{B}_c=1$ and the point of beyond quasiparticle quantum non-locality.}
    \label{fig:checkfrozen}
    \end{figure}

It should be noted that freezing the middle site was simply a choice made for numerical simulations; freezing any other site would yield qualitatively similar results. This implies that to obtain the plot showing results for different distances, one does not need to repeat step (1) for each distance. Instead, one can repeat step (2) by decoupling different sites at varying distances each time to obtain the variation of Liang information with distance. This is verified in Fig.~\ref{fig:checkfrozen}, showing that similar qualitative results can be achieved by keeping the target site fixed and varying the position of the frozen site, which is operationally simpler.
\end{document}